\documentclass[preprint,12pt]{elsarticle}
\makeatletter 
\def\ps@pprintTitle{%
 \let\@oddhead\@empty
 \let\@evenhead\@empty
 \def\@oddfoot{}%
 \let\@evenfoot\@oddfoot}
\makeatother

\usepackage{amsmath,amssymb,amsfonts}
\usepackage{algorithmic}
\usepackage{graphicx}
\usepackage{textcomp}
\usepackage{float}
\usepackage{subfig}
\usepackage{xcolor}
\usepackage{hyperref}
\usepackage{balance}

\journal{}
\hypersetup{
pdftitle={Measurement and characterization of DNS over HTTPS traffic},
pdfsubject={DNS over HTTPS},
pdfauthor={Kamil Jeřábek, Ondřej Ryšavý, Ivana Burgetová},
pdfkeywords={Domain Name System, DNS over TLS, DNS over HTTPS, DoH, Privacy and Security, Network Traffic Measurement},
}

\begin{document}

\begin{frontmatter}



\title{Measurement and characterization of \\ DNS over HTTPS traffic}


\author[inst1]{Kamil Jeřábek}

\affiliation[inst1]{organization={Faculty of Information Technology, Brno University of Technology},
            addressline={Bozetechova 1/2}, 
            city={Brno},
            postcode={61200}, 
            country={Czech Republic}}

\author[inst1]{Ondřej Ryšavý}
\author[inst1]{Ivana Burgetová}

\begin{abstract}
Domain name system communication may provide sensitive information on users' Internet activity. DNS-over-TLS and DNS-over-HTTPS are proposals aiming at increasing the privacy of Internet end users. In this paper we present an overview of the current state in the deployment of DNS-over-HTTPS (DoH) implementations complemented by measurements of DoH traffic in terms of the incurred overhead and the possibility of the DoH detection based on the inherent characteristics of the communication patterns.
\end{abstract}




\end{frontmatter}


%
%
\section{Introduction}
\label{sec:introduction}
Internet these days is a witness of efforts to increase the security and privacy of its users. As of 2021, about 85\% of all Internet communication is protected by encryption\footnote{https://www.fortinet.com/blog/industry-trends/keeping-up-with-performance-demands-of-encrypted-web-traffic}. On the other hand, the most of domain queries on the Internet is plain text communication, which reveals a lot of information about user interactions even before users access the service. Technically, Internet Service Providers (ISP), enterprise network administrators, and others with access to the Internet traffic can easily observe and analyze the DNS content, thus violating users' privacy.

Moreover, DNS without any other protection is also vulnerable to various attacks. Attempts to secure the entire DNS environment include, in particular, Domain Name System Security Extension (DNSSEC), which delivers missing integrity and authenticity verification of the DNS records\cite{rfc2535}. The DNSSEC was introduced two decades ago, but its adoption rate is very slow\cite{chung2017understanding}. From the privacy perspective, this extension does not encrypt the DNS traffic itself, and thus anyone can still inspect the content of DNS queries.

In general, proposed methods to protect DNS traffic encapsulate requests and responses into other protocols such as TLS, DTLS, HTTPS. However, apart from the fact that those solutions encrypt the traffic, it also brings the overhead in terms of increased payload, extra packets, and initialization delay, leading to additional latency. For each proposal, it is thus important to understand the costs of security and privacy in terms of traffic overhead. 

The contribution of this paper consists of the following:

\begin{itemize}

\item We overview features of domain name resolution protocols: DNS, DNS over TLS, and DNS over HTTPS. The option to use the resolution using those protocols is implemented in new versions of the popular web browsers (Firefox, Chrome). 

\item We analyze existing protocol implementations of DoH clients in web browsers and DoH server deployments. In particular, we performed several experiments aiming to compare the protocol performance characteristics, either isolated or in the expected application scenarios.

\item We created a large dataset of DoH communication in terms of captured packets. By analysis of traffic samples, we evaluate DoH performance (see Sections V and VII) and characterize DoH traffic features, which illustrates how DoH communication differs from other HTTPS traffic (see Section VIII). Also, we provide an explanation of the use of HTTPS headers in DoH communication. The provided insight can be, for instance, used for developing efficient methods of DoH traffic classification.
\end{itemize}

The paper is organized as follows: The next section overviews relevant protocols. Section III discuss the current state of DoH implementation in client-side applications. Section IV presents a novel tool for detecting and testing deployed DoH servers on the Internet. It also provides results on the experiments that identified various parameters currently available servers. Section V consists of the analysis of performance parameters of considered domain resolution protocols in isolated settings. Section VI focuses on the study of DoH implementations in web browsers. Section VII describes experiments measuring web page load times, showing the practical impact of DoH. In Section VIII the characterization of DoH traffic is provided together with a comparison of DoH and regular web traffic. Finally, the paper is concluded in Section IX by discussing the results.
%
%
\section{Background}
This section overviews the principles of i) the original DNS as privacy extensions are built on top of this protocol, ii) HTTP that provides a transport mechanism for DNS over HTTPS, and iii) DNS over HTTPS (DoH), which seems to be a major domain resolution protocol of the future privacy preserving domain name system. 

\subsection{Domain Name System}
\begin{figure}[h]
    \centering
    \includegraphics[scale=0.59]{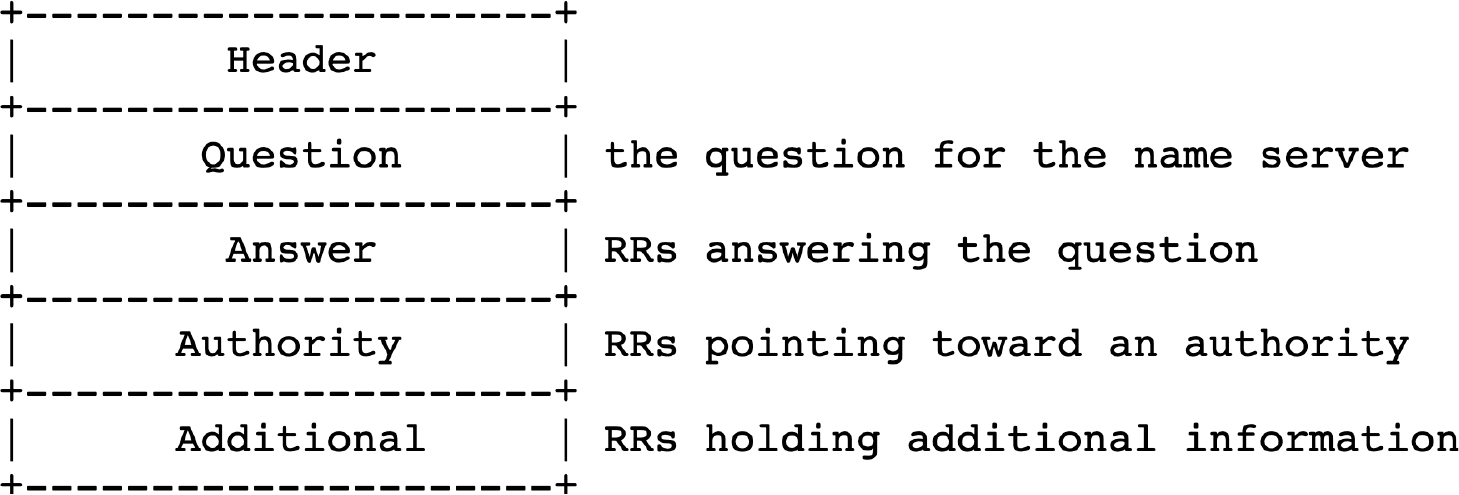}
    \caption{DNS message format.\cite{rfc1035}}
    \label{fig:dns-message-format}
\end{figure}

The Domain Name System provides a DNS protocol that enables hosts the translation of domain names into IP addresses. The protocol was first introduced in RFC 1035\cite{rfc1035} and extended by numerous RFCs later\footnote{https://help.dyn.com/articles/dns-rfcs/}. The communication uses messages of a simple format is depicted in Figure \ref{fig:dns-message-format}.

The message is divided into multiple parts, where some of them are optional. The message always contains a header determining whether it is a question or an answer. The header is followed by the question part that carries three fields: query type, query class, and query domain name. This part is followed by a possibly empty list of resource records with the \texttt{Answer} that answers the question, the \texttt{Authority} that points toward an authoritative name server, \texttt{Additional} section contains additional information related to the query but not strictly answering the question itself\cite{rfc1035}.

The DNS message can be transmitted using both UDP and TCP transport. The reserved port of DNS is 53 in both cases. The UDP payload messages are restricted to 512 bytes. TC bit is set in the header for a longer DNS message, and the message is truncated. In the case of TCP protocol, the two-byte length field is added before the DNS message in the TCP payload, giving the length of the message. A DNS message can use the maximum TCP payload size, and it is not limited to only 512 bytes, as in UDP.




\subsection{HTTP(S)}
The HTTP/1.0 was followed by the HTTP/1.1 that introduced extensions to the protocol and improvements such as persistent connections, advanced caching options, compression, and so forth\cite{rfc2616}. 

The HTTP/2 was introduced in 2015 by RFC 7540. The protocol version supports all the core features but overcomes the previous versions' inefficiency and extends the functionality that correlates better with today's needs. The major difference is that it enables prioritization and multiplexing, which allows multiple streams packed within the single connection\cite{rfc7540}. The protocol is more complicated to implement than the previous versions.

To secure HTTP communication, the messages are transmitted in TLS provided encrypted channel. HTTP embedded in TLS is denoted as HTTPS. The TLS 1.3 \cite{rfc8446} is the newest version currently available, but an abundance of HTTPS traffic use TLS1.2. 

\subsection{DNS over HTTPS}
DNS over HTTPS is a fairly new protocol defined in RFC 8484\cite{rfc8484}. DNS queries and responses are transferred using HTTPS. Thus HTTPS is a transport protocol for them. In principle, a DNS query is represented either as HTTP POST or GET method. The DNS servers are required to support both options. 

\subsubsection{POST Method}
\begin{figure}[h]
    \centering
    \includegraphics[scale=0.97]{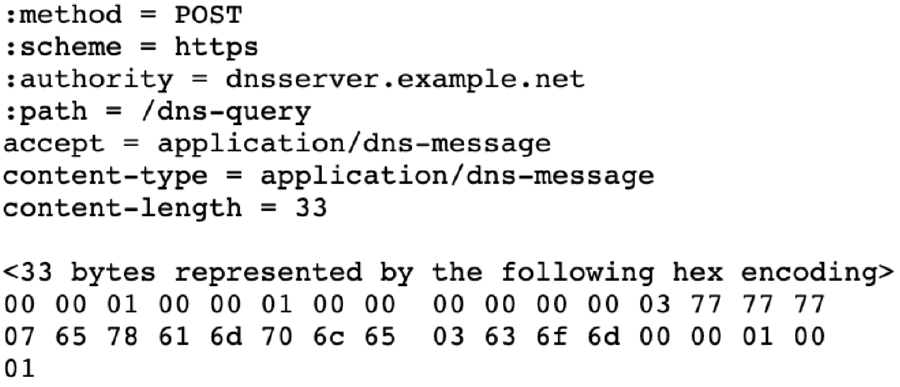}
    \caption{DOH POST query example \cite{rfc8484}}
    \label{fig:doh-post-query}
\end{figure}

Figure \ref{fig:doh-post-query} depicts POST DNS example query. The HTTP header contains request field \texttt{Content-Type} representing media type of value \\\texttt{application/dns-message}. The body of the message contains a DNS query in the DNS wire format\cite{rfc1035}. The HTTP header may contain \texttt{Accept} field to clarify what type of response the client expects, e.g., binary or JSON.

\subsubsection{GET Method}
\begin{figure}[h]
    \centering
    \includegraphics[scale=0.95]{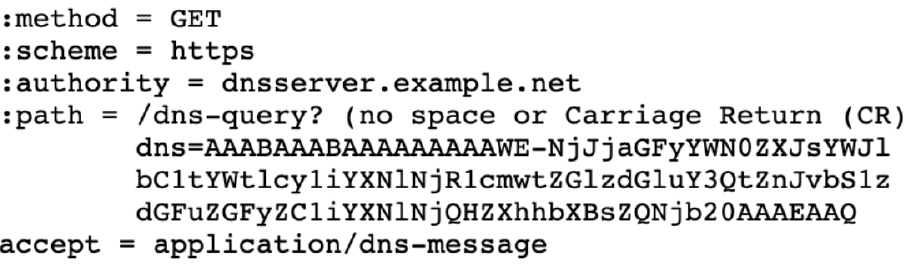}
    \caption{DOH GET query example \cite{rfc8484}}
    \label{fig:doh-get-query}
\end{figure}

The DNS query using HTTP GET is shown in figure \ref{fig:doh-get-query}. The body of the request is empty as the queried domain is encoded in \textit{base64url} format as the value of the \texttt{dns} variable.

\subsubsection{DNS Response}
\begin{figure}[h]
    \centering
    \includegraphics[scale=0.95]{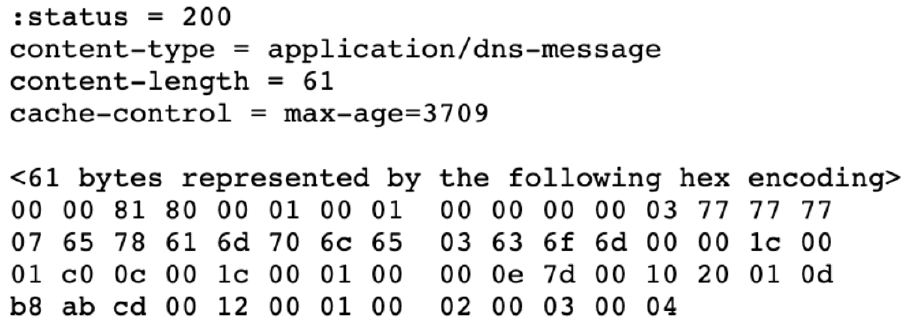}
    \caption{DOH response example \cite{rfc8484}}
    \label{fig:doh-response}
\end{figure}

The DNS response (see example in Fig. \ref{fig:doh-response}) should be of type \\\texttt{application/dns-message}, though RFC 8484 \cite{rfc8484} notes that 
other formats may be introduced. The data payload for this media type is a single message in the on-the-wire format. The maximum DNS message size is defined to be 65535 bytes. 

One DNS request and response should be mapped to one HTTPS message exchange, but in HTTP/2, the multistreaming functionality can be involved for the replies.

\begin{figure}[h]
    \centering
    \includegraphics[scale=1]{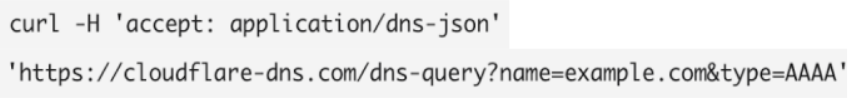}
    \caption{DOH JSON request example}
    \label{fig:doh-json-request}
\end{figure}

\begin{figure}[h]
    \centering
    \includegraphics[scale=1.1]{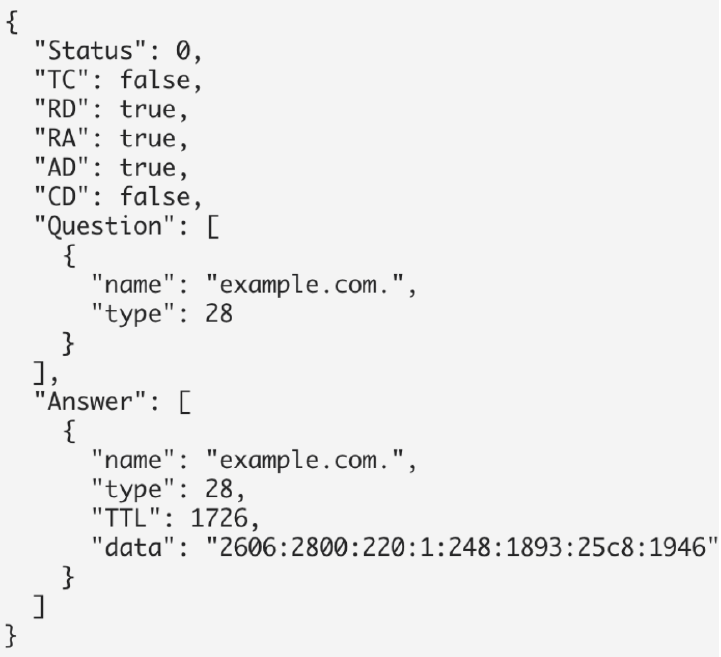}
    \caption{DOH JSON response}
    \label{fig:doh-json-response}
\end{figure}

Global DNS resolver providers such as Cloudflare and Google also implement the opportunity to exchange DNS information using JSON format. This format is not included in the RFC standard. In this case, the client set \texttt{content-type} to \texttt{application/dns-json}.
The request URI can be seen in Figure \ref{fig:doh-json-request}. The main variables are \texttt{name} for the domain name in plain text and \texttt{type} for the DNS response.
The response can be seen in Figure \ref{fig:doh-json-response}. The response fields and information is similar for both Google and Cloud Flare providers (provided examples come from Cloud Flare developers documentation\cite{cloudflare-json-documentation}).

%
%
\section{Current State}
The DNS over HTTPS is the fairly new protocol defined a few years ago and drafted in RFC 8484 by P. Hofmann (from ICANN) and P. McManus (from Mozilla) \cite{data-tracker-doh-draft}. The publication of this RFC unleashed a wave of interest as well as criticisms.

The high effort of content delivery network providers in combination with application and operating system vendors accelerated the adoption of the DoH protocol. Mozilla added the first support for DoH in version 62 of their Firefox browser \cite{firefox-first-doh-version-62}, and in February 2020 DoH was the default option for all US users \cite{firefox-us-users}. Chromium project introduced support for DoH in version 83 of Chrome browser \cite{chrome-first-doh-version-83}. Support in other browsers based on Chromium code-base such as new Edge, Brave, Opera, followed shortly. Furthermore, Chrome for Android devices introduced the support since version 85 \cite{chrome-first-android-doh-version-85}. Microsoft enables DoH in Windows 10 operation system insider preview build 19628 by 13th of May 2020 \cite{microsoft-doh-first-build}. The DoH was also announced by Apple for iOS version 14 and macOS version 11\cite{apples-doh-support}. However, the Apple Safari browser does not support DoH yet.

Recently, the number of DoT/DoH servers has increased significantly as Chaoyin Lu, et. al. \cite{lu-end-to-end-large-scale-measurement-doe} reveals. 
The migration from DNS towards DoH has an impact on the web browsing in different host environments and providers as identified in series of works by Austin Hanousel, et. al.\cite{hounsel-analyzing-costs-of-dns-dot-doh, hounsel-comparing-effects-of-dns-dot-doh} shows. Their measurements reveal that the DoH outperforms DoT in the case of longer queries. Traditional DNS has a better response time than DoH and DoT mainly because of the overhead introduced by encrypted underlying transport channels of DoH and DoT.




%
%
\section{DoH servers}

DoH servers and DoH clients are the primary building blocks of the new DNS privacy-enabled environment. For conducting the presented experiments, we need to have access to a reasonably significant sample of different existing DoH deployments. Therefore, in this section, we briefly overview the current state of public DoH server deployment and features of major DoH server implementations. 

Lu, et. al. \cite{lu-end-to-end-large-scale-measurement-doe} claimed that it is very difficult to discover DoH servers on the Internet. Hence, the index of publicly available DoH servers \cite{curl-doh-list} maintained by the Curl community is the primary information source. The list contains providers with their known DoH servers. Community, as well as large CDN providers, are both included in this list. Unfortunately, the list is maintained manually providing only a brief description of each endpoint. To improve the situation we created a monitoring tool specifically focused on testing DoH servers that used information from this list. The tool periodically checks DoH server features, including their availability. This tool is available online\cite{doh-sigman}.

\begin{figure}[h]
    \centering
    \includegraphics[scale=0.58]{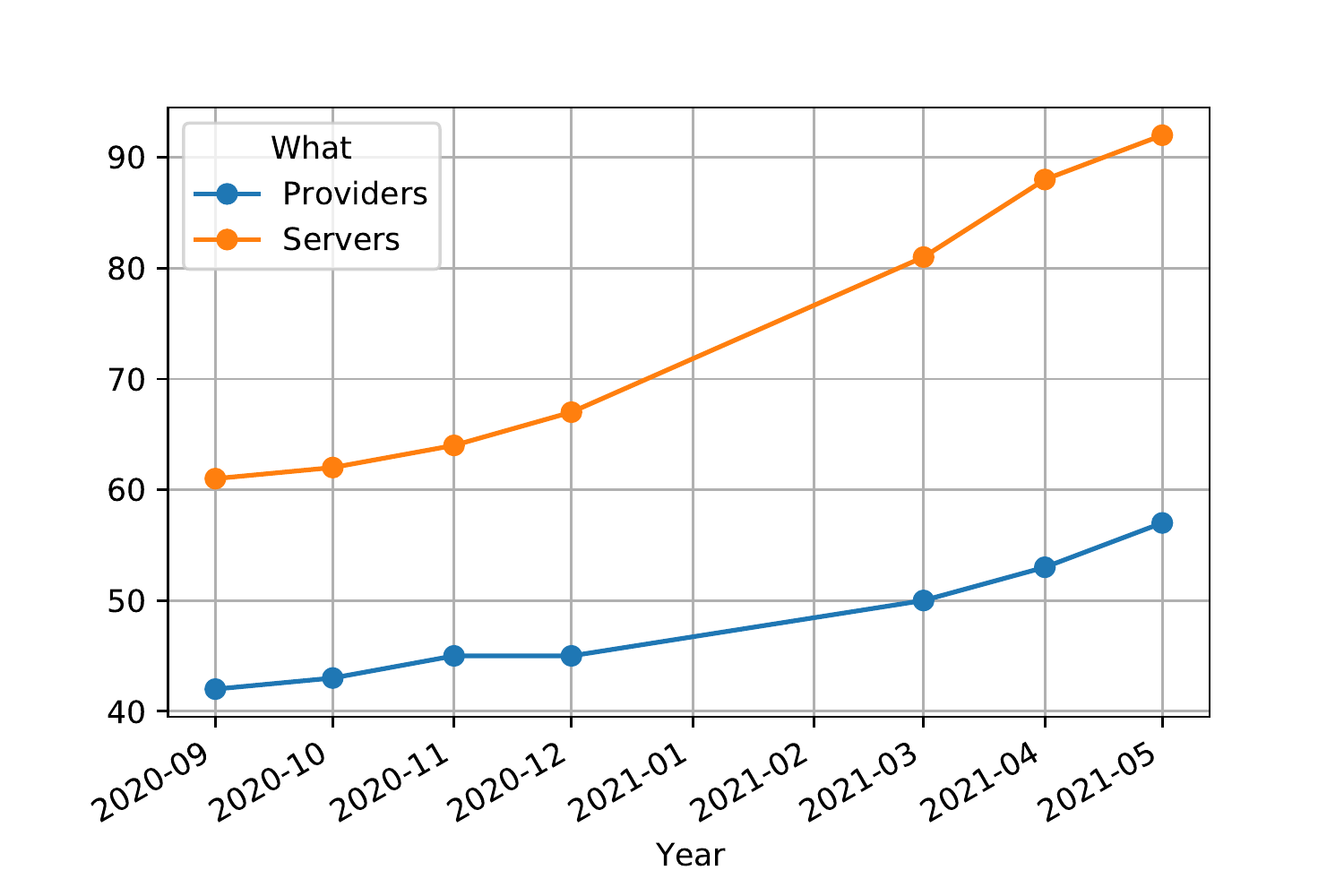}
    \caption{DOH providers and servers.}
    \label{fig:providers_servers}
\end{figure}

The tool has been running since September 2020\footnote{Note that in January and February 2021 the server running the tool was under maintenance thus values from this period are missing.} providing statistics as shown in Fig. \ref{fig:providers_servers}). The DoH servers that do not correctly respond to requests were removed. From the presented data, it can be observed that some providers ($17$) offer multiple DoH endpoints. Also, it can be seen that the numbers of providers and servers raise.

\begin{table}[ht]
\begin{center}
        \begin{tabular}{|l|c|c|} \hline
            \textbf{Stats} & \textbf{Counts} & \textbf{Percentage} \\ \hline
Servers & 92	& \\ \hline
HTTP2 &	88 &	95.7 \% \\
TLS1 &	19 &	20.7 \% \\
TSL1.1 &	20 &	21.7 \% \\
TSL1.2 &	91 &	98.9 \% \\
TLS1.3 &	81 &	88.0 \% \\
IPv4 &	90 &	97.8 \% \\
IPv6 &	73 &	79.3 \% \\
Let's Encrypt &	54 &	58.7 \% \\ \hline
        \end{tabular}
\end{center}
    \caption{Currently available DOH servers stats. Data obtained from \cite{doh-sigman}.}
    \label{tab:doh-servers-stats}
\end{table}

Each live DoH server was periodically contacted to test various parameters: TLS versions, HTTP versions, IP versions, supported DoH requests, certificates installed (see Table \ref{tab:doh-servers-stats}). The presence of the Let's Encrypt certificate signalizes the installation of experimental or small communities servers. The recent data shows that almost half of the servers are of this type. From the total of 92 observed Doh servers,  $95.7\%$ of servers run on HTTP/2 which is recommended by RFC 8484. These servers also commonly support HTTP/1.1. Nevertheless, a small number of DoH servers run only on HTTP/1.1 which is considered experimental and not recommended for production use. However, in our study, we consider both HTTP protocols. Another important observation is that some servers use deprecated TLS versions (<1.2). Deployment of servers with deprecated TLS may lead to known security risks \cite{rfc8996}. Finally, the DoH servers vary in the support of the IP versions. 

The preliminary observations presented in this section give us valuable information on differences in DoH deployments that we consider in the experiments conducted as a part of our study and presented in forthcoming sections.
%
%
\section{Single DoH Query Analysis}

The first experiment aims at measuring single and separated DoH requests. This gives us an estimation of the expected overhead of DoH compared to the traditional DNS transaction. 
Also, results are used to set the baseline for the identification of unsuccessful DoH requests generated by browsers as used in Section \ref{sec:browseranalysis}.

The RFC 8484 \cite{rfc8484} defines that the DoH supports two HTTP methods to be used for querying: GET or POST. In addition to those two methods, an experimental DoH with JSON encoding is supported by big DNS providers such as Cloudflare, Google, etc. Thus JSON encoded DoH is considered as another variant in our experiments. 
We generated and analyzed the same set of queries for all four request methods: DNS, DoH using GET, DoH using POST methods, and DoH with JSON.

\subsection{Data Generation}
The domain names used in queries stem from the top one million domains from the Cisco Umbrella dataset\cite{cisco-umbrella-dataset}. Only resolvable domains were included in the dataset, yielding approximately $800,000$ domain queries for each type of request method. All queries were done towards Cloudflare servers.

There are only a few standalone tools able to perform just a single query. Those tools are command-line tools such as nslookup (for DNS queries), Curl community developed simple curl-doh (for DoH), and we developed our simple command-line tool for querying DoH for both GET and POST commands. Using browsers and proxies is unsuitable because we cannot control the number of queries in a single flow.

\begin{figure}[H]
    \centering
    \includegraphics[scale=1.2]{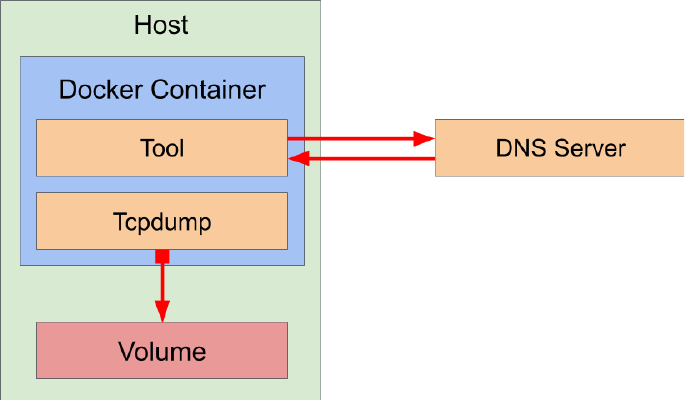}
    \caption{Container structure for data generation.}
    \label{fig:singlerequests-dataset-capture-structure}
\end{figure}

To run our toolset, we deploy a Docker container that enables us to capture network traffic on its interface (see Fig. \ref{fig:singlerequests-dataset-capture-structure}). The hardware platform was Supermicro SuperTwin2 6026TT-TF server equipped with eight Intel (R) Xeon E5520 @ 2.26 GHz. The cluster consists of 4 nodes. The nodes are equipped with 48 GB RAM and 16 CPU cores. The cluster is connected via 1Gb/s links to the campus network. The tools run simultaneously on all cluster nodes. The dataset is available at\cite{doh-my-dataset}.

\subsection{Traffic Analysis}

The captured packets were preprocessed by grouping them into flows to compute the following statistics: payload size, number of packets, and overall flow duration. As flow duration follows the normal positively skewed distribution, we used medians for comparison. Fig.9 consists of graphs showing the overhead of DoT and DoH in comparison with the traditional DNS.

\begin{figure*}[t]
    \subfloat[Median payload \\size of flows comparison.]      {
        \includegraphics[width=.33\linewidth]{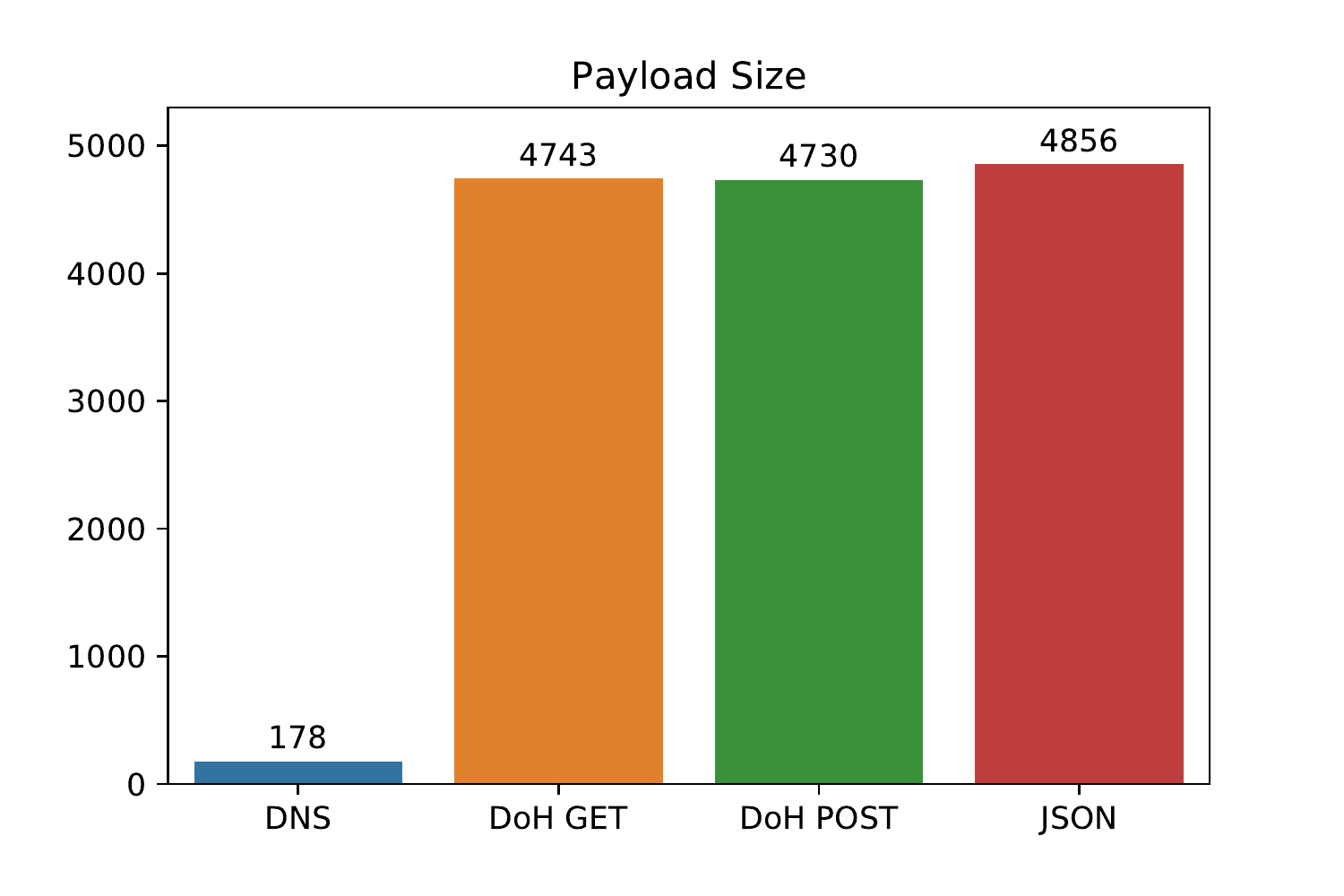}
        \label{fig:comparison-single-flows-paysize}
    }
    \subfloat[Median packets \\in flows comparison.]      {
        \includegraphics[width=.33\linewidth]{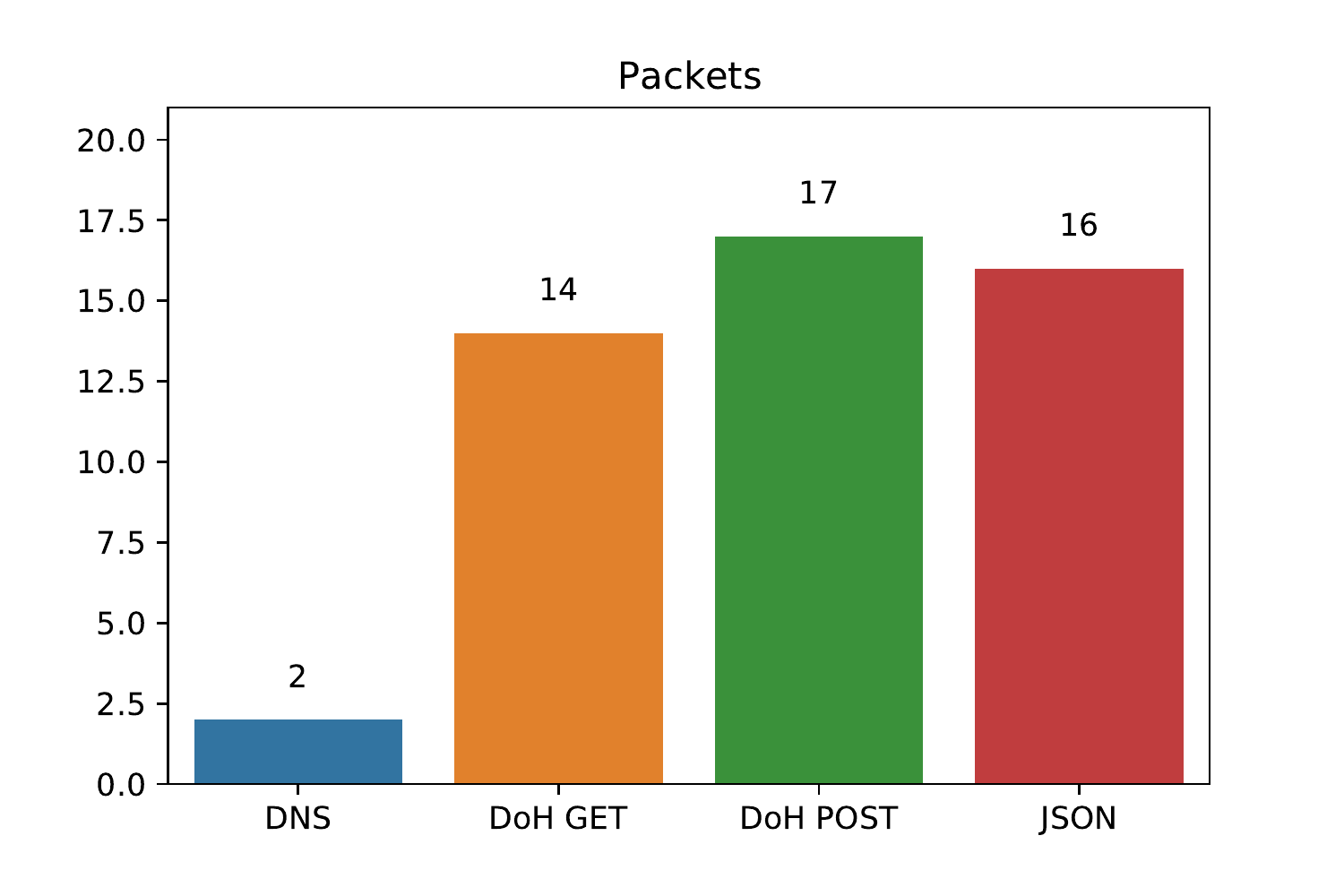}
        \label{fig:comparison-single-flows-packets}
    }
    \subfloat[Median duration \\of flows comparison.]      {
        \includegraphics[width=.33\linewidth]{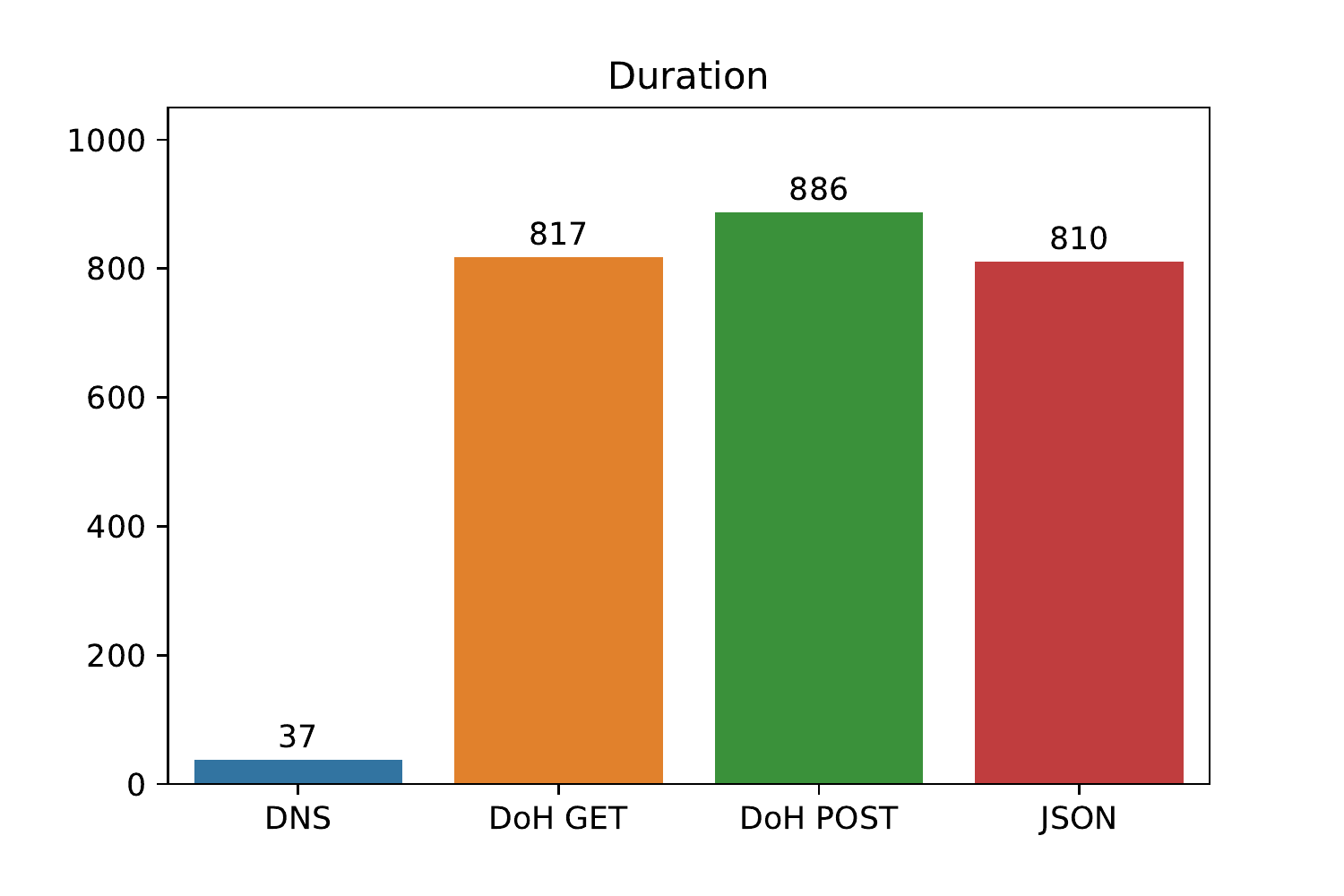}
        \label{fig:comparison-single-flows-duration}
    }
    \\
    \subfloat[Median payload \\size of flows comparison.]      {
        \includegraphics[width=.33\linewidth]{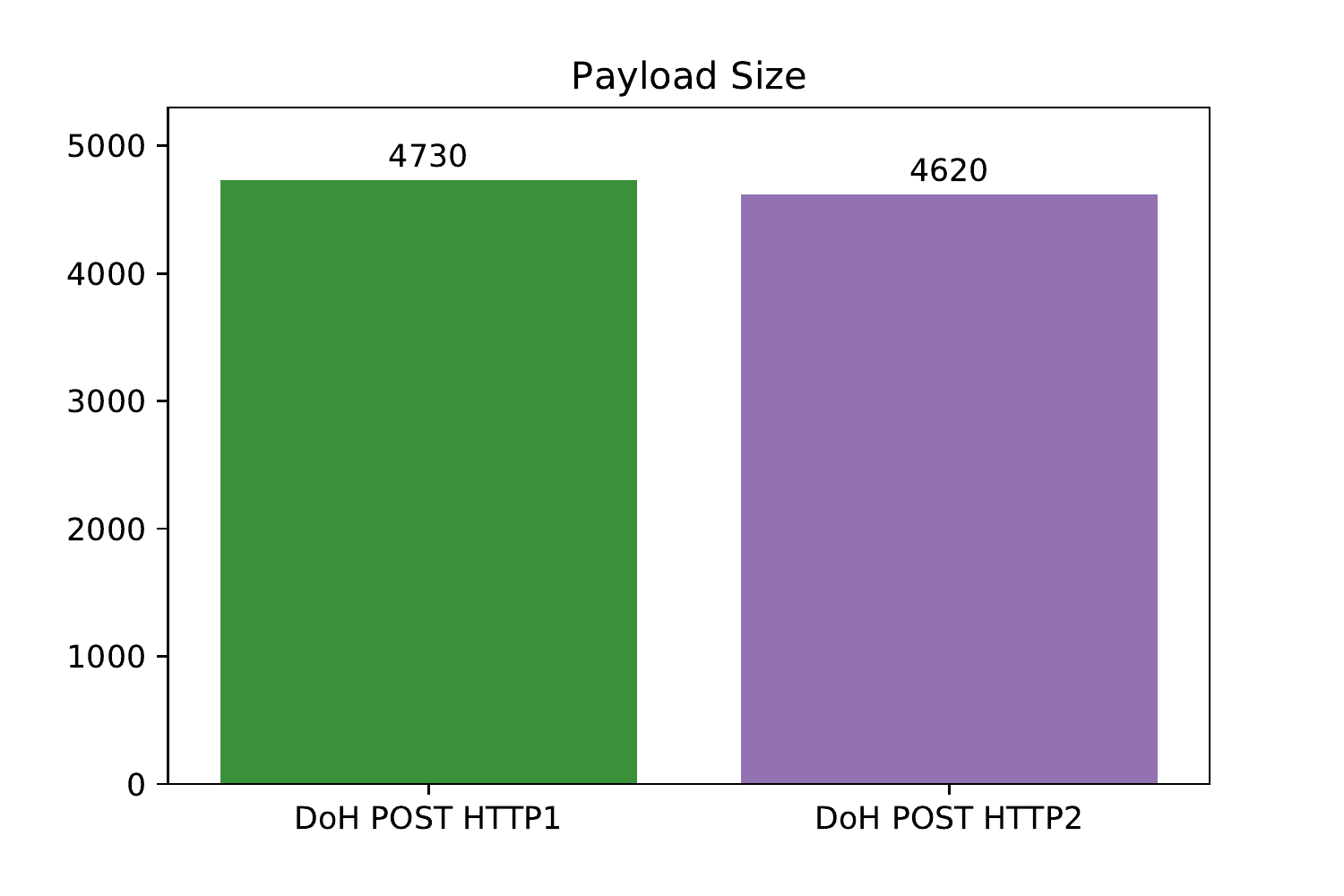}
        \label{fig:comparison-single-flows-http-payload}
    }
    \subfloat[Median packets \\in flows comparison.]      {
        \includegraphics[width=.33\linewidth]{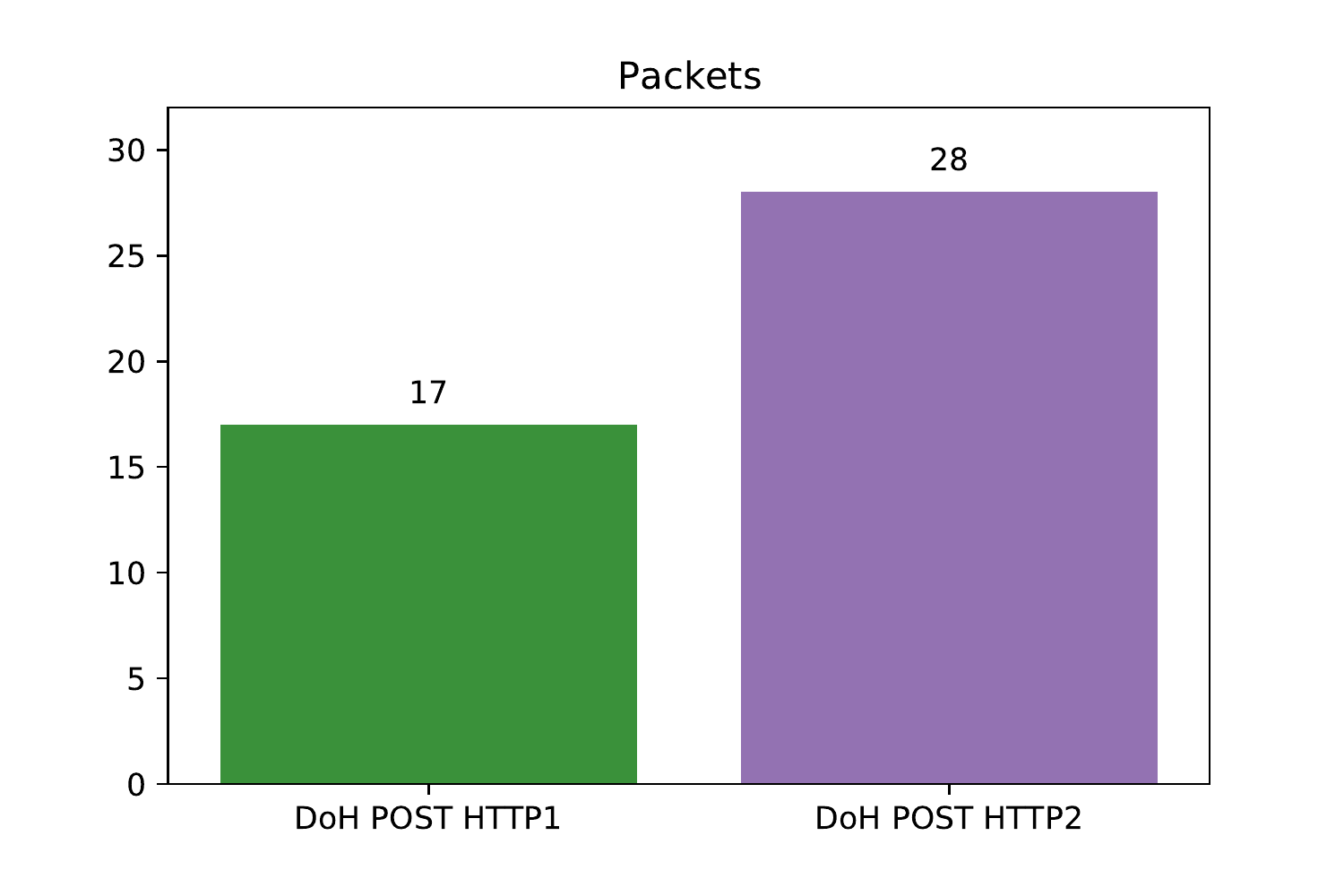}
        \label{fig:comparison-single-flows-http-packets}
    }
    \subfloat[Median duration \\of flows comparison.]      {
        \includegraphics[width=.33\linewidth]{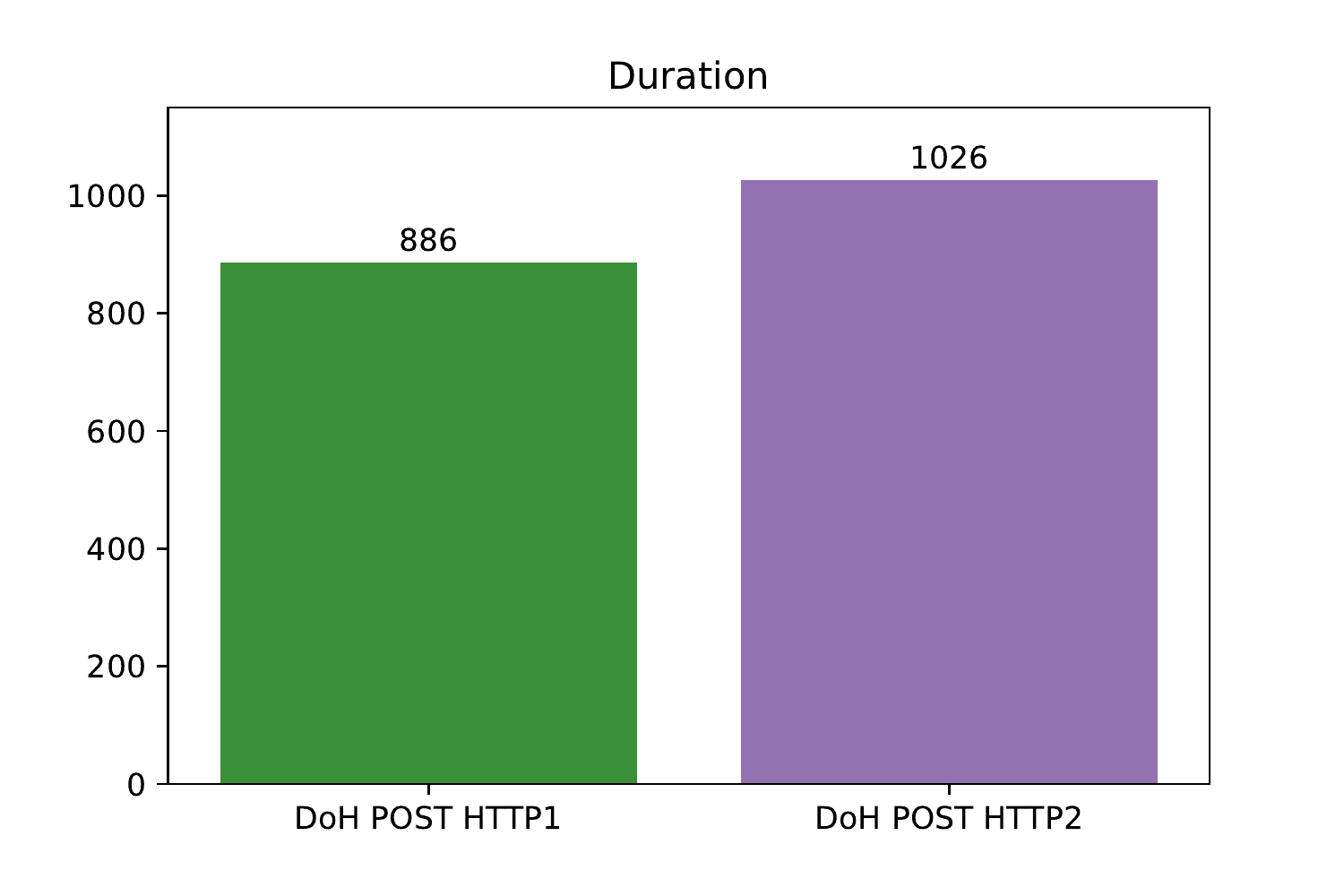}
        \label{fig:comparison-single-flows-http-duration}
    }
    \caption{Comparison of DNS and DoH options statistics.}
    \label{fig:comparison-single-flows-all}
\end{figure*}

Figure \ref{fig:comparison-single-flows-paysize} shows the medians of the payload sizes of the whole flows for single pairs of DNS queries. To transmit a single DNS query and response of about $178$ bytes using clear DNS, we need nearly $27$ times more bytes to transmit DoH or DoH with JSON. The difference between GET and  POST, and JSON methods in terms of payload sizes of the whole flow is not that significant. 
Similarly, there is also a significant difference in terms of packets (see Fig. \ref{fig:comparison-single-flows-packets}) and the overall duration of the flow (see Fig. \ref{fig:comparison-single-flows-duration}).

Thus, for individual DNS queries, DoH has significantly worse every performance parameter than original DNS communication. It is mainly caused by the fact that only a fraction of the DoH flow is occupied by DoH request or response data, while the rest is mainly the overhead of TLS handshake. Of course, this overhead is reduced when a TLS connection is utilized for more DNS queries as done, for instance,  by most DoH clients. 

On the other hand, the differences between individual DoH and DNS over JSON are not so significant and each method exceeds the other methods in one of the monitored statistics. In terms of payload size, the most efficient encrypted method is DoH with POST method. The GET method shows a $0.27\%$ increase in payload size over the POST method, DNS over JSON shows $2.66\%$ increase (based on median values). From the perspective of the number of packets, the most effective method is DoH with GET method. POST method utilizes $21.43\%$ more packets, DNS over JSON uses $14.29\%$ more packets in average.
The DNS over JSON shows the shortest duration from the encrypted methods. HTTP method GET requires $0.86\%$ more time, method POST $9.38\%$ more time compared to DNS over JSON.

In addition, we also measured queries using HTTP/2 to compare the behavior of the DoH using HTTP/1.1. We used the POST method to show the difference between using those two versions of HTTP. A slight difference can be observed mainly in the number of packets (see Fig. \ref{fig:comparison-single-flows-http-packets}) as HTTP/2 frames and splits the message into separate packets, which also leads to a bit longer duration of the HTTP/2 flows (see Fig. \ref{fig:comparison-single-flows-http-duration}). However, both are almost the same in terms of overall flow size (see Fig. \ref{fig:comparison-single-flows-http-payload}).
%
%
\section{DoH-enabled Browser Analysis}
\label{sec:browseranalysis}

Besides the experimental implementations of DoH clients in OSes, the web browsers are the most common applications currently utilizing DoH. Thus, it no longer holds that domain resolution is the service offered by the operating system involving DNS communication. If enabled, the browser actively communicates with a DoH server to translate domain names to the corresponding server addresses needed to establish HTTPS connections with them. The web browser has to know an appropriate DoH server address. In 2021, the DoH feature can be enabled in most currently available browsers.
The configuration varies depending on the browser. For instance, browsers based on the Chromium code-base have few predefined and hard-coded servers for enabling the DoH, together with the possibility to set user-provided DoH server. Firefox provides default DoH, which is Cloudflare's resolver, but a user can manually insert any other DoH server. 

This section contains an analysis of DoH clients, represented by two selected web browsers.  We chose browsers primarily because they provide a stable implementation of DoH clients. Many users often unintentionally use DoH running browser as they have DoH enabled by default. We used web browsers to generate a dataset of DoH communication for our further analysis. 

\subsection{Chrome}
Chrome browser provides an implementation for DoH since version 78 (only experimental at that time). In the experimental version, the DoH was only activated if the system's DNS servers were set to one of the servers in a hard-coded table. Chrome released official support for DoH in version 83. Currently, the number of provided hard-coded servers\cite{chromium-hardcoded-table} is increasing, and users can set their own server when needed. The setting of Chrome DoH is tightly related to the system settings. Chrome still does not support DoH on Linux, but it has supported Android OS since version 85.

DoH in Chrome works as follows: 
i) Chrome DoH client asks the system configured DNS servers to resolve the following \verb|dns.google| domain. 
ii) If a valid server address is retrieved, it is used to establish the HTTPS connection to the DoH server.
iii)  The DoH session is used for resolving all other domain names. The DNS message is Base64 encoded and sent as a query parameter in HTTP/2 GET command. The answer is sent in a data stream of HTTP/2 in the DNS wire format. 

Chrome contains a fallback mechanism if the DoH service replies error code. In this case, Chrome tries to resolve the domain using the system's DNS resolver. Chrome DoH also supports EDNS padding \cite{rfc8467} which allows DoH clients and servers to increase the size of a DNS message, thwarting the size-based correlation of encrypted DNS messages.

\subsection{Firefox}
Firefox has provided a more tuneable environment for DoH since the beginning of the support of the DoH protocol. Users can set modes of DoH (force, try, off), methods used for resolution (GET, POST), and the particular DoH server. The default provider is Cloudflare, but it is possible to specify any custom provider. 
When using the predefined provider, the browser attempts to resolve its domain name and selects one of the retrieved IP addresses to establish a DoH connection. In default settings, it queries \verb|mozilla.cloudflare-dns.com| domain.
Upon successful acquisition of a DoH server address,  the browser starts multiple connections with the server. One of the connections lasts for the entire lifetime of the web browser session. This connection is kept alive by using \verb|PING| commands of HTTP/2 if necessary. The \verb|POST| method sends DNS queries in the wire format. In the answer, the data stream also contains DNS wire format, which is indicated by  \verb|application/dns-message| content type. The requests coming from browser applications are uniform, either GET or POST. The change that can be observed is whether the server uses HTTP/2 or HTTP/1.1. When writing this paper, the Firefox browser does not support EDNS padding as Chrome does. It is currently reported as an enhancement in Mozilla's bug report system but not solved\cite{edns-bug-firefox}.

\subsection{Data Generation}
\begin{figure}[h]
    \centering
    \includegraphics[scale=0.84]{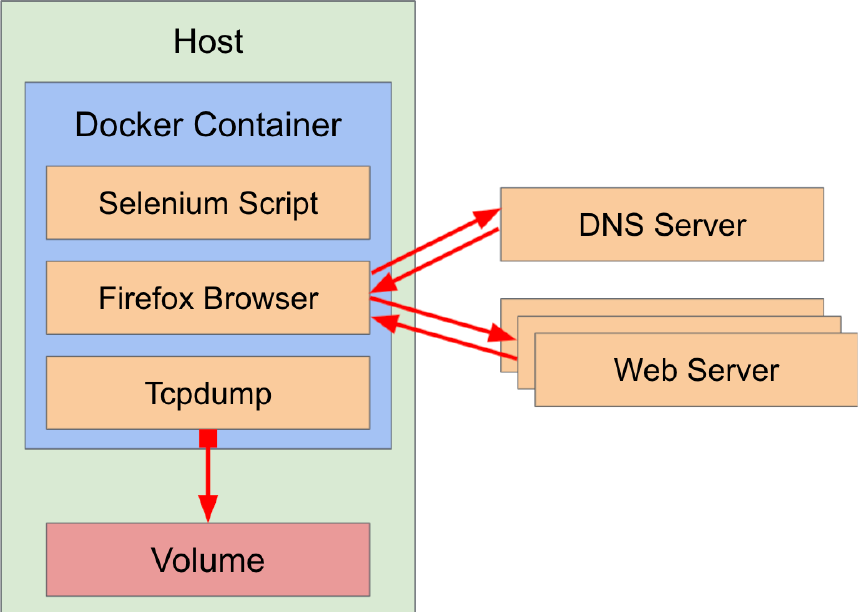}
    \caption{DoH generation browser.}
    \label{fig:schema-browser-doh-generation}
\end{figure}

We used web browser engines to generate DoH datasets for further analysis. The domains for webpages downloaded were taken from the Majestic Million dataset\cite{majestic-million-dataset} which is suitable for direct access to existing web pages. The primary browser chosen to generate data is Firefox because of the configuration variability compared to Chrome. The data are generated by browsers running in the Docker containers as depicted in Figure \ref{fig:schema-browser-doh-generation}. The traffic is then captured on the Docker network interface.

The browser does not run in headless mode. Instead, X virtual frame buffer is used. It implements the X11 display server protocol where all graphical operations are made in virtual memory without graphical output\cite{xvfb}. This mimics the browser's usual usage of the browser by a user. The Selenium script drives the browser operations.

The browser runs with all cache disabled and DNS cache expiration set to 0, which requires that every domain be resolved and not get from the cache. The requests for the web pages are made separately. Before each request, the browser is opened, and after each request, the browser is closed. Using this operational pattern, the browser generates a higher amount of DoH traffic. The generated data are part of the whole dataset that is available at\cite{doh-my-dataset}.

%
%
\section{Browser Page Load Impact}

In this section, we present another experiment that aims at measuring the impact of DoH on web applications. Specifically, we measure and compare web page load times if name resolution is made using DNS and DoH, respectively. 

We employ the same infrastructure as in the previous experiment. The measurements run simultaneously on separate machines. About $10,000$ pages were loaded entirely in the web browser. We use the same Cloudflare's resolver for domain resolution using DNS, DoH GET, and DoH POST methods. The measurement was repeated three times to avoid outliers possibly caused by unexpected network congestions. The round trip times of the resolver for both DoH and DNS were similar. The round trip time was measured, and the average difference between DNS and DoH servers from our measurement point was $0.02ms$. Cloudflare's DoH resolver accepts HTTP/2 connections.
\begin{figure}[h]
    \centering
    \includegraphics[scale=0.54]{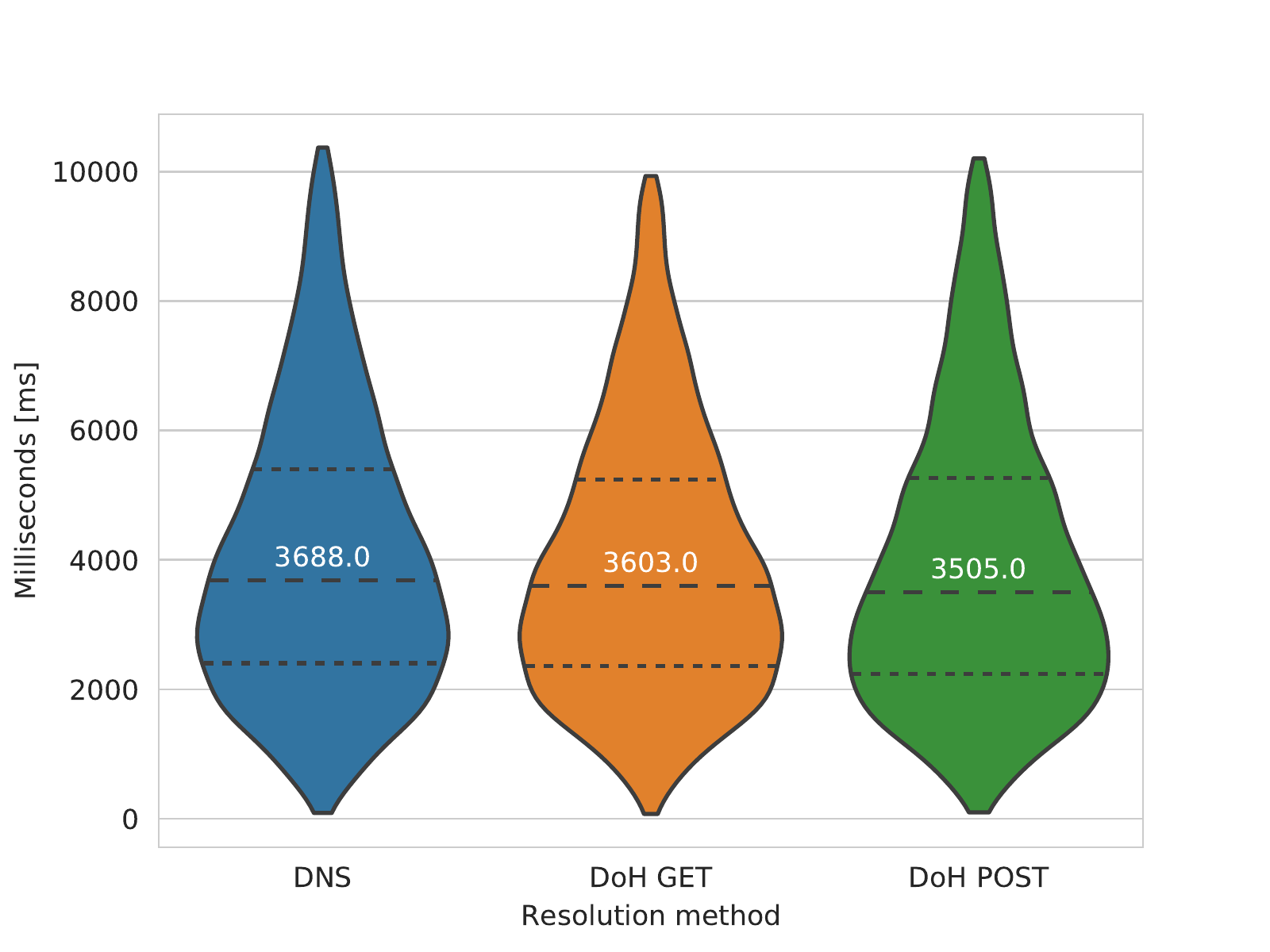}
    \caption{DoH page load time comparison of DNS, DoH using GET and POST towards Cloudflares server.}
    \label{fig:browser-time-measurements}
\end{figure}
Figure \ref{fig:browser-time-measurements} displays a graph that represents measurement of page load times. Values in the graph represents differences between \texttt{domainsLookupStart} and \texttt{domComplete} browser events.

The distribution of the differences in all categories has a similar shape. According to the positively skewed distribution, the medians were taken as the leading value in comparison. Surprisingly, the DNS method appears to be the slowest, while and DoH using the POST method is the fastest, with a difference of about $\sim100ms$ between DoH methods and $\sim200ms$ between slowest DNS and fastest DoH POST. The difference between DNS and DoH POST according to our measurements reaches $\sim5\%$. The difference is relatively low, but we can consider that it is present here.

\begin{figure}[h]
    \centering
    \includegraphics[scale=0.54]{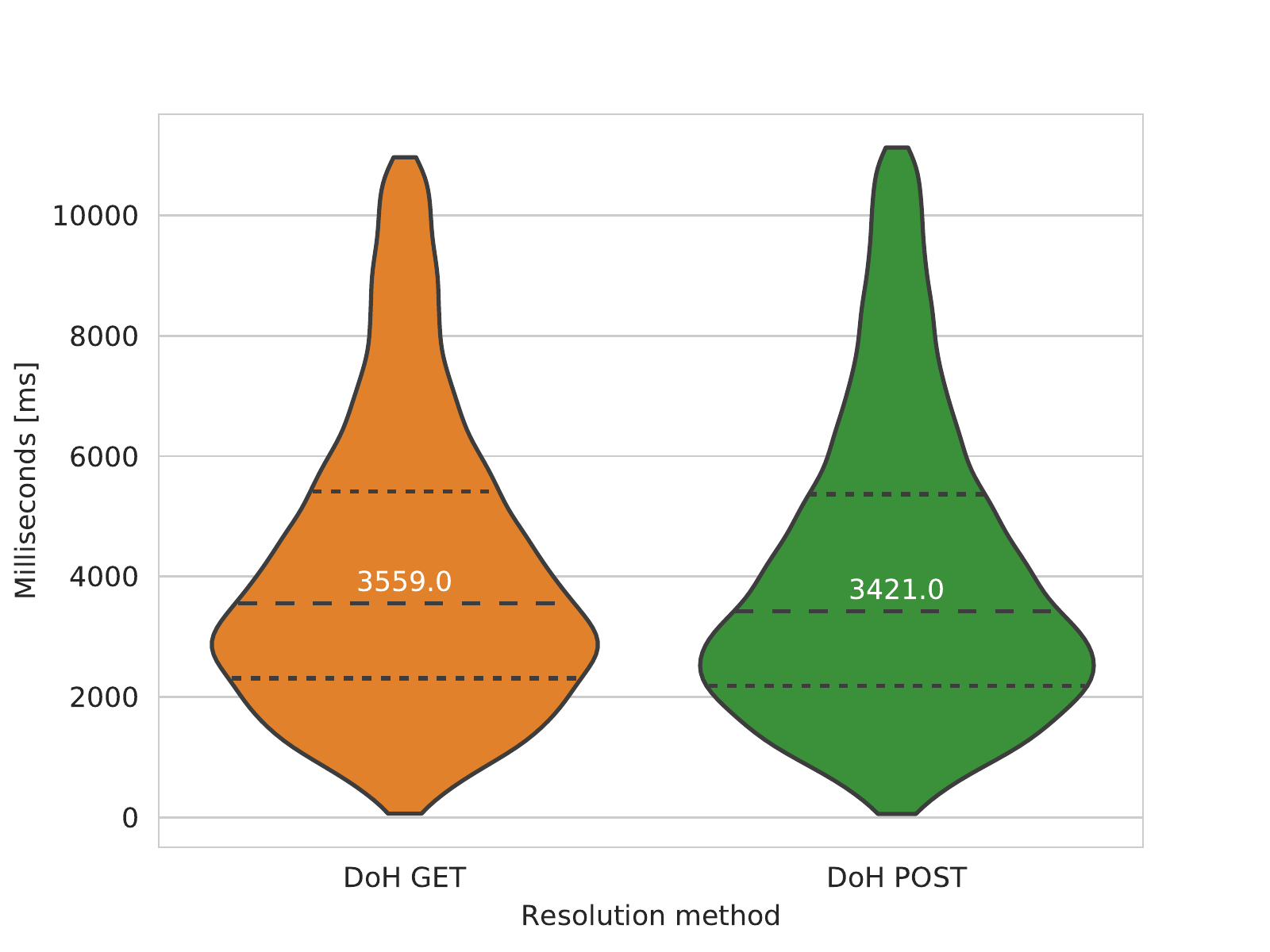}
    \caption{DoH page load time comparison DoH using GET and POST towards Google server.}
    \label{fig:browser-time-measurements_google}
\end{figure}

To confirm this finding, we compared both DoH methods in the experiment involving Google's DoH resolver. The source of domains was taken from the same dataset as previously, except for the number of domains reduced to $30,000$ pages. Figure \ref{fig:browser-time-measurements_google} depicts the comparison, and we can observe a similar gap between medians of resolution methods as in the previous measurement. In both cases, the DoH POST method has slightly better performance in page load times.

The POST method is a winner in this experiment with respect to web page load latency. Also, POST requests are generally smaller than their GET equivalents. However, this is because, in our test, we do not consider the impact of the domain name caching mechanism. In reality, domain name caching can reduce the time needed for resolution, and the GET method may be more suitable as the responses to POST requests are not cacheable in general. Privacy Google's best practices for DoH \cite{google-doh-privacy-best-practices} claim that using the POST method is more suitable for privacy critical applications, where caching is undesirable.

The previous section reveals DNS as the fastest option for single queries. It is a different result than in the case of the impact of using the resolving protocols on page load times. Here, we should consider that the overhead caused by TCP and TLS handshakes at the beginning of the DoH flows is amortized over the flows carrying multiple requests.


\section{Browser Requests Multiple Servers}

The previous experiments reveal the minimal impact on page load time between the DoH methods. In this section, we analyze a characteristic of DoH communication considering different DoH servers. Many providers currently deploy DoH servers using various DoH software, versions, configuration, operating systems, etc., in their installations. Moreover, the DoH resolvers are part of the HTTPS ecosystem, including load-balancers, reverse proxies, etc. Thus to determine the real-world DoH traffic characteristics, we need to consider an entire system. We generate and analyze the DoH traffic using the same environment as in previous experiments, considering the same list of domains, but we target a sample of available DoH servers with different deployment features. We divided the servers into two broad groups depending on whether HTTP/1.1 or HTTP/2 is used as the transport protocol. This experiment aims to characterize DoH traffic by showing the similarities and differences of individual types of servers. Also, we present how DoH traffic differs from other HTTPS communication. To represent it, we used simple but very representative metrics consisting of the ratio of the number of packets and average packet size for DoH connections. 

\begin{figure}[h]
    \centering
    \includegraphics[scale=0.35]{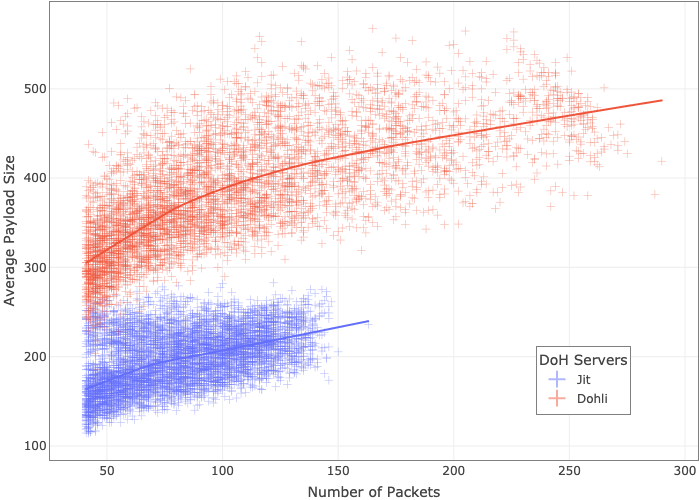}
    \caption{HTTP/1.1 DoH servers payload size to packets comparison}
    \label{fig:http1-doh-servers-packets-paysize}
\end{figure}

Only a small portion of the servers exclusively use HTTP/1.1. We chose two of them and showed the ratio of average TCP payload sizes to packet numbers (we include packets with zero data, e.g., empty TCP acknowledgment packets). This is shown in Figure \ref{fig:http1-doh-servers-packets-paysize}.
The observable difference is mainly because of HTTP headers in DoH responses generated by the server. While web browser (DoH client) always sends similar HTTP headers. Each server sends headers containing various HTTP headers, including unnecessary ones, e.g., \texttt{X-Powered-By}. The HTTP/1.1 does not compress the headers sent in plain text, representing a significant overhead in each message. The size of header lines in most cases exceeds the amount of DNS data. DNS resolution should be fast. Applications (primarily browsers) sends multiple requests in a short period. The head-of-line (HOL) blocking should cause a considerable performance decrease in the case of HTTP/1.1. The browsers are trying to overcome this problem by opening multiple flows and sending requests in parallel. Unfortunately, the problem with HOL blocking is still present there.
 
\begin{figure}[h]
    \centering
    \includegraphics[scale=0.35]{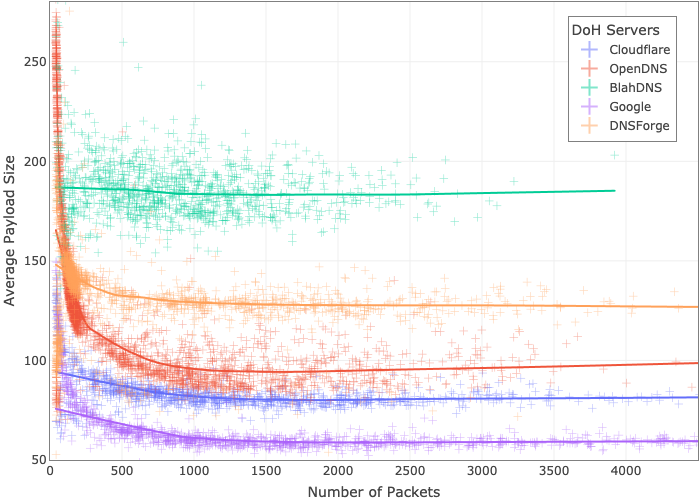}
    \caption{HTTP/2 DoH servers payload size to packets comparison.}
    \label{fig:http2-doh-servers-packets-paysize}
\end{figure}

The second group of servers primarily exposes DoH on HTTP/2. The majority of the servers from the DoH list support HTTP/2. Again Figure \ref{fig:http2-doh-servers-packets-paysize} depicts the ratio of packet size to packet number in this case for HTTP/2 DoH servers.
It can be observed that the average size of the packets is lower than in the case of HTTP/1.1, which is caused by the design of HTTP/2. The HTTP/2 performs HPACK header compression \cite{rfc7541} in each request/response. Moreover, data and headers can be transferred separately, which lowers the average payload size and increases the number of packets. One HTTP/2 flow can hold multiple streams, leading to an increased number of packets exchanged within a single flow. Interestingly, using this characterization, it is possible to distinguish different DoH servers.

Note the difference in the shape of curves for HTTP/1.1 and HTTP/2 cases. The flow always begins with a TLS handshake that has a similar characteristic for all communications that significantly contributes to the size/packet ratio. The TLS handshake is amortized near $1,000$ packets in a flow. Thus for long-term DoH connections, corresponding curves converge to their specific constant values.

\begin{figure}[h]
    \centering
    \includegraphics[scale=0.35]{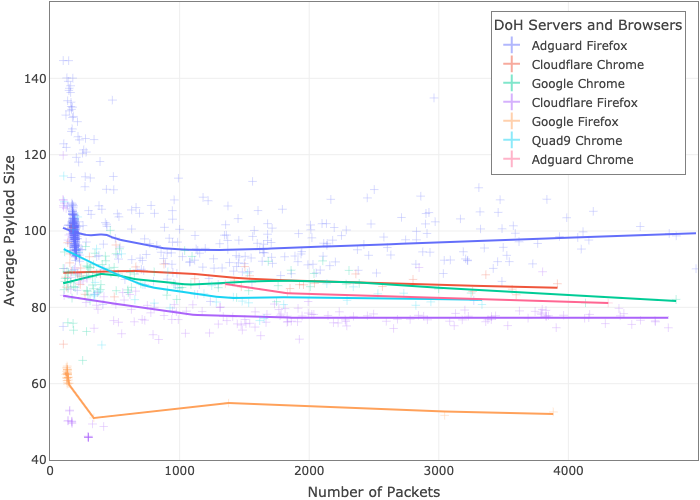}
    \caption{CIRA-CIC-DoHBrw-2020 dataset HTTP/2 DoH servers payload size to packets comparison.}
    \label{fig:cicd-doh-servers-packets-paysize}
\end{figure}

To get more reliable results, we also determine the same characteristics for publicly available DoH traffic CIRA-CIC-DoHBrw-2020 dataset\cite{cicd-article,cicd-dataset-page}. The results are shown in Figure \ref{fig:cicd-doh-servers-packets-paysize}. According to the dataset specification, data were generated from the Firefox and Chrome DoH communication with servers on HTTP/2. The dataset has fewer flows than ours but with a longer duration. Nevertheless, characteristics similar to ours can be observed. The difference between multiple servers using the same browser, as well as the difference between using Chrome and Firefox browsers towards the same servers, can be identified, which confirms our observation. Moreover, it is shown that the Firefox browser has a lower ratio than Chrome when using the same servers. It is because of Chrome's implementation of EDNS padding.

Interestingly, DoH network traffic of different web browsers and DoH servers can be distinguished by using a simple payload size to packet number ratio. The protocol analysis shows that EDNS extension and HTTP header size are the most significant contribution to this aspect.

\begin{figure}[h]
    \centering
    \includegraphics[scale=0.35]{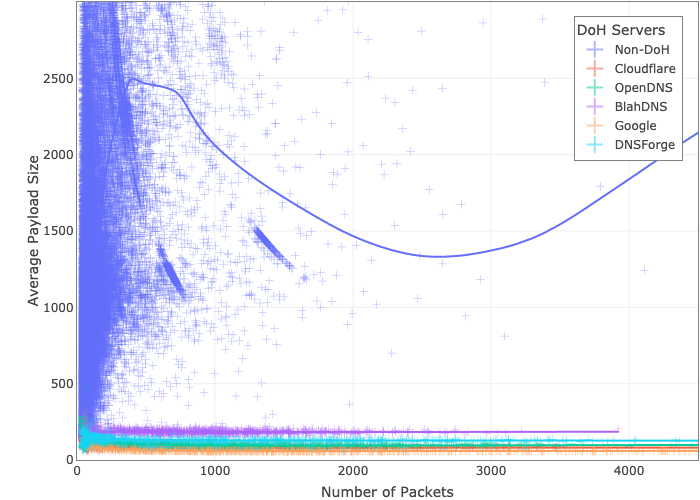}
    \caption{HTTP/2 DoH servers payload size to packets comparison together with the rest non-DoH HTTPS traffic.}
    \label{fig:http2-doh-servers-packets-paysize-normal}
\end{figure}

Despite the observations of only DoH traffic, we can also observe non-DoH traffic present in both datasets. Figure \ref{fig:http2-doh-servers-packets-paysize-normal} depicts both DoH traffic from multiple DoH servers and the non-DoH traffic using blue color. The traffic is generated specifically as described earlier, and separated requests towards domains from the list were made for a short period. Hence, the non-DoH traffic lacks longer streams and interactions, as can be observed from the graph. We can see mostly flows carrying images, short presentation videos, and other standard documents. The depicted non-DoH traffic contains larger packets (reaching up to 64 KB), which is caused by enabled TCP offloading in virtualized infrastructure. This fact does not change the validity of DoH characteristics, but provides view of non-DoH traffic in one of the possible environments. The graph is zoomed to be able to see the difference between DoH and non-DoH traffic rather than the full picture of non-DoH traffic.

\begin{figure}[h]
    \centering
    \includegraphics[scale=0.35]{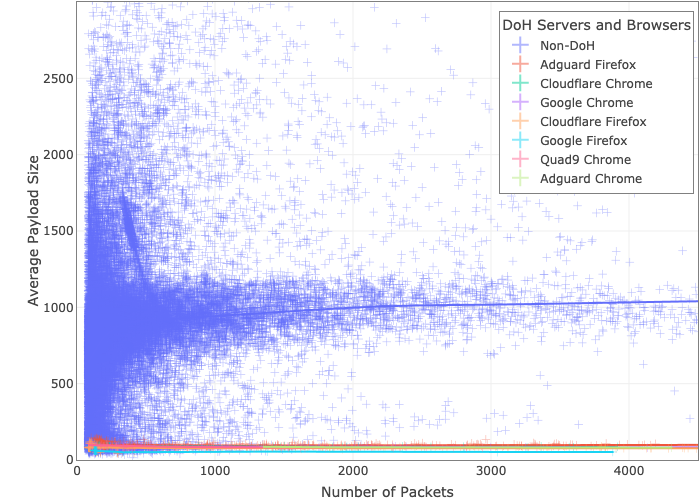}
    \caption{CIRA-CIC-DoHBrw-2020 dataset HTTP/2 DoH servers payload size to packets comparison together with the rest of non-DoH HTTPS traffic.}
    \label{fig:cicd-plus-normal}
\end{figure}

Figure \ref{fig:cicd-plus-normal} depicts similar observation, except it comes from the CIRA-CIC-DoHBrw-2020 dataset. The dataset is generated differently. Hence the characteristics of the non-DoH traffic are different. During the generation, longer interactions were made, and the non-DoH traffic gets higher diversity and behaved more realistically. However, it can be seen that the DoH can still be distinguished considering the feature of payload size to packet number ratio.

\subsection{Headers}
HTTP headers have a direct impact on the amount of DoH traffic.\\ HTTP/1.1 DoH communication is affected more than HTTP/2 because of plain text headers\footnote{Servers operating only over HTTP/1.1 are somewhat experimental or temporal. Also, the portion of such servers in the DoH server list is very small (about $4.3\%$)}. 
The HPACK header compression may partially help. Nevertheless, to reduce this overhead, it is necessary to remove unnecessary header lines carefully.

\begin{figure}[h]
    \centering
    \includegraphics[scale=0.58]{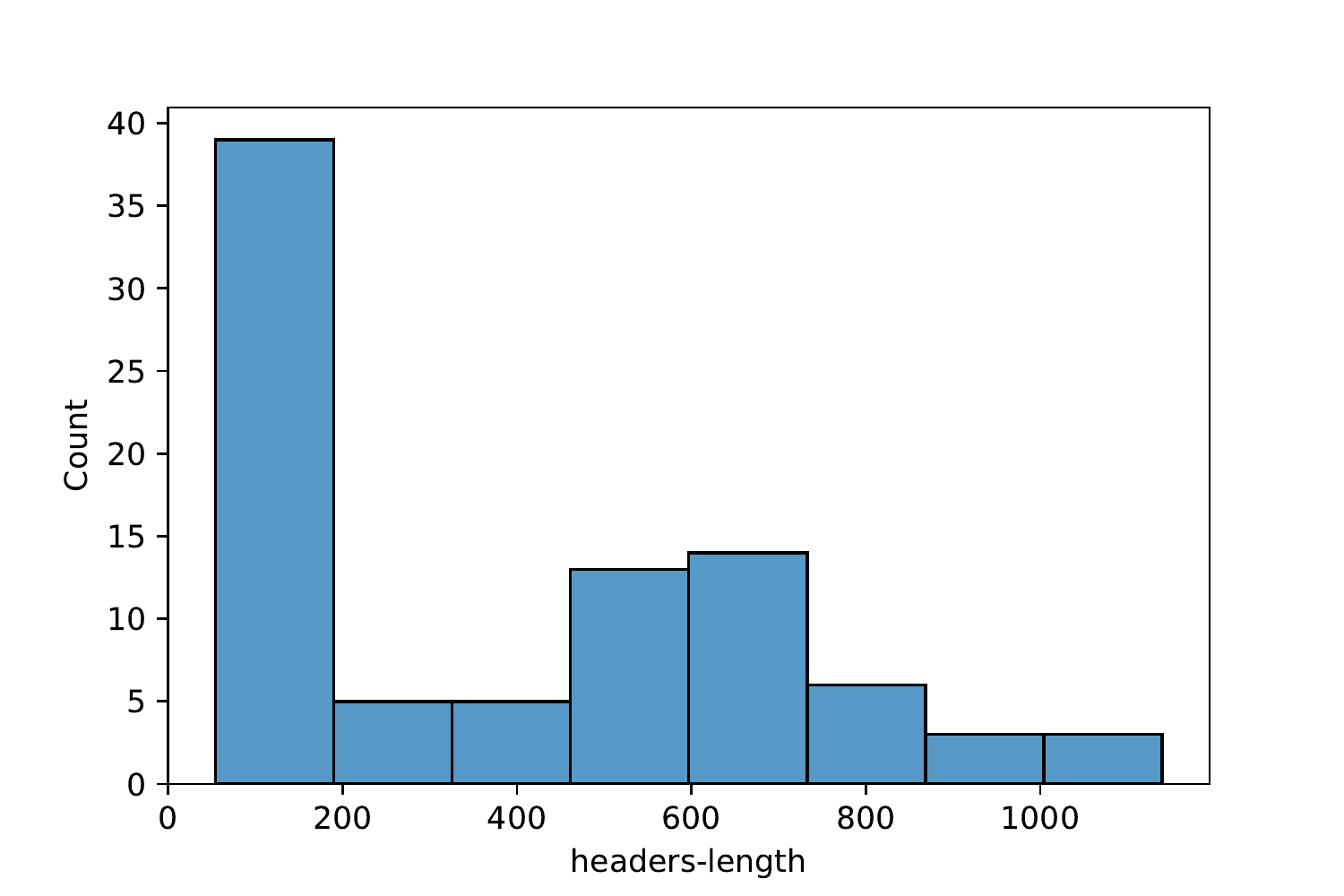}
    \caption{Headers length histogram.}
    \label{fig:headers-length-histogram}
\end{figure}

To determine the current situation, we observe the header that the servers put to the responses (see Fig. \ref{fig:headers-length-histogram}). It can be seen that about half of the servers use header conservatively with an entire size of fewer than $200$ bytes. However, a considerable number of servers produce response headers of larger sizes, some of around $1,000$ bytes, which is a significant overhead. 

\begin{figure*}[h]
    \centering
    \includegraphics[width=.9\linewidth]{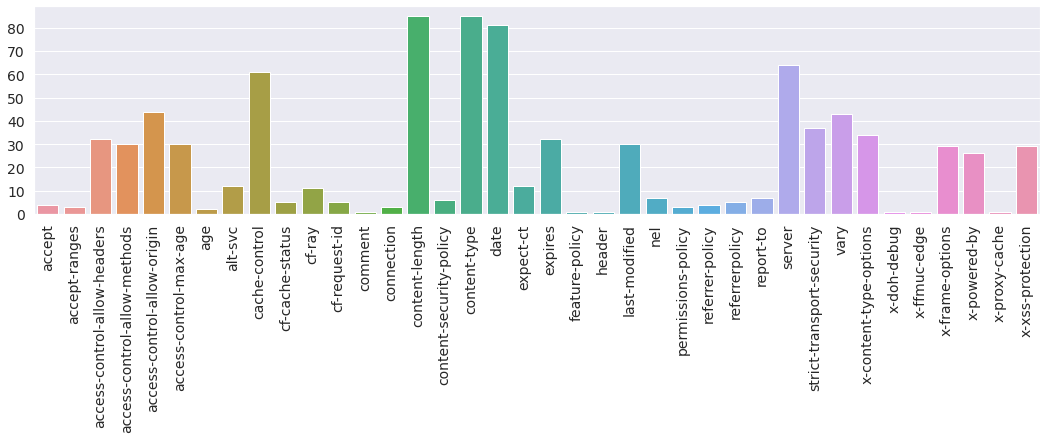}
    \caption{Frequency of DoH options in HTTP headers}
    \label{fig:headers-cetnost-full}
\end{figure*}

The detailed analysis is presented in Fig. \ref{fig:headers-cetnost-full} that depicts the individual observed header items in DoH server responses. Items \texttt{content-type} and \texttt{content-length} are required and thus they were found in all responses. The other header items are optional and vary greatly in observed responses. More than half of servers are using \texttt{cache-control} option to express the proper cache operations for replies. According to our measurements, only two do not follow RFC 8484 properly.

The header often also identifies server software and its configuration. Header \texttt{x-powered-by}, when provided, reveals an implementation of DNS over HTTPS server software ($23$ items)\footnote{DNS over HTTPS client, server software available on https://github.com/m13253/dns-over-https}. The \texttt{server} header reveals servers running on some version of Nginx ($34$ items), Cloudflare CDN ($12$ items), and  h2o/dnsdist ($11$ items).


%
%
\section{Conclusion}
DNS over HTTPS is an experimental protocol that aims to increase the privacy of Internet users by providing an encrypted communication channel for DNS message exchange. It is not the first attempt at adding security to the DNS. Still, it seems that its support by application vendors and Internet service providers gives this protocol a chance of wide adoption, in particular, compared to its short-lived predecessor called DNS over TLS. In this study, we identified the currently available DoH server, showed the growth of a number of running instances, and detected supported DoH protocol features. Next, we performed a simple experiment to demonstrate the overhead of DoH by generating a dataset of single DNS queries towards Cloudflare's servers. We compared the performance of original DNS, DoH POST, DoH GET, and DoH with JSON encoding. 

In a more complex experiment, we aim at performance measurement when the DoH is used in its target environment. We generated data from the web browser to demonstrate the possible differences in page load times for different DNS resolving methods. In this case, the DoH POST can be considered a slightly faster option beating even the traditional DNS method.

By observing network traffic, we found no significant differences between DoH POST and GET methods if communicating with the same DoH server. However, differences can be spotted for network traffics communicating with different servers depending on whether the servers use HTTP/1 or preferred HTTP/2. Another notable feature that impacts the amount of transferred data is the size of HTTP headers. Therefore the DoH clients and servers should restrict the size of HTTP headers as it may be greater than the total size of DNS data transferred.

Finally, we provided a characterization of DoH traffic using accurate but straightforward metrics. We showed that although various DoH deployments have slightly different metrics, the DoH traffic characteristics are significantly different from the other HTTPS traffic. Thus, it is possible to develop accurate methods of DoH traffic identification as already presented in the literature \cite{vekshin2020, 9312004, essay82085} but also to identify the different software implementations of DoH servers and their deployments.

\section*{Acknowledgment}
This work was supported by Brno University of Technolgy, Faculty of Information Technology internal grant FIT-S-20-6293.

\bibliographystyle{elsarticle-num} 
\bibliography{main}





\end{document}